# Anisotropic Strain Engineering in La$_{0.7}$Sr$_{0.3}$MnO$_3$/LaFeO$_3$ Superlattice: Structural Relaxation and Domain Formation

**Authors**:

*Yu Liu, Thea Marie Dale, Emma van der Minne, Susanne Boucher, Romar Avila, Christoph Klewe, Gertjan Koster, Magnus Nord, Mari-Ann Einarsrud, Ingrid Hallsteinsen[*]*

**Affiliations:**
Y. Liu, M.-A. Einarsrud, I. Hallsteinsen
Department of Materials Science and Engineering, NTNU Norwegian University of Science and Technology, Trondheim, Norway
*corresponding author Email: ingrid.hallsteinsen@ntnu.no

T.M. Dale, S. Boucher, M. Nord
Department of Physics, NTNU Norwegian University of Science and Technology, Trondheim, Norway

E.v.d. Minne, R. Avila, G. Koster
MESA+ Institute for Nanotechnology, University of Twente, Enschede, The Netherlands

C. Klewe
Advanced Light Source, Lawerence Berkeley National Laboratory, Berkeley CA, US

Funding: The work is internally funded by NTNU and The COST Action CA-20116, European Network for Innovative and Advanced Epitaxy (OPERA) Short-Term Scientific Missions (STSMs). The Research Council of Norway is acknowledged for the support to the Norwegian Micro- and Nano-Fabrication Facility, NorFab, project number 295864, Norwegian Center for Transmission Electron Microscopy, NORTEM (197405) and In-situ Correlated Nanoscale Imaging of Magnetic Fields in Functional Materials, InCoMa (315475).

Keywords: anisotropic strain engineering, transition metal oxide thin film, strain relaxation, structural twin domain

## Abstract

Anisotropic strain engineering in epitaxial oxide films provides new opportunities to control the antiferromagnetic and structural properties crucial for advancements of antiferromagnetic spintronics. Here we report on a (La$_{0.7}$Sr$_{0.3}$MnO$_3$/LaFeO$_3$)$_4$ superlattice grown on (101)$_o$ DyScO$_3$ substrate which imposes significant anisotropic in-plane strain. Reciprocal space mapping reveals selective strain relaxation along the tensile in-plane [010]$_o$ axis, while compression along the perpendicular in-plane [$\bar{1}$01]$_o$ axis remains strained. Scanning precession electron diffraction and higher-order Laue zone analysis show that the relaxation is accommodated by structural domain formation in the LaFeO$_3$ layers, initiating from the second bilayer and propagating out-of-plane. These domains minimise structural defects and correlate with the substrate step edges. X-ray magnetic dichroism measurements reveal bulk-like in-plane antiferromagnetic order with polydomain signature as previously reported. Our findings reveal the presence of structural domains coexisting with antiferromagnetic polydomain states, showing a strain-domain-magnetism relationship that provides insights for applications of strain engineering in spintronics applications.

## 1. Introduction

Antiferromagnetic (AF) materials are rapidly emerging as key candidates for next-generation spintronic technologies. [1, 2] Unlike ferromagnets, AF materials are spin-compensated with magnetic sublattice, giving no macroscopic magnetisation. This intrinsic spin compensation eliminating stray field and enabling ultrafast spin dynamics, offers major advantages for energy-efficient and high-speed devices. However, the absence of macroscopic magnetic moment makes the spin orientation difficult to control. In addition, the lattice imperfections and crystal twinning often results in AF multidomain formation. Thus, AF materials have primarily been used as passive components such as pinning layers for the active ferromagnets for exchange bias applications. [3, 4]

Complex oxide perovskites, such as $LaFeO_3$ (LFO) and $La_{0.7}Sr_{0.3}MnO_3$ (LSMO), exhibit crystal structures composed of corner-sharing $BO_6$ octahedra that are highly responsive to epitaxial strain. LFO is a G-type antiferromagnet in which strain-induced distortions modify the magnetocrystalline anisotropy and enable control over the Néel vector. [5, 6] LSMO, a double-exchange ferromagnet, similarly exhibits strain-tunable magnetic anisotropy [7, 8], making both materials well suited for studying strain–spin coupling in perovskite thin films. By choosing a low-symmetry substrate, the AF domain degeneracy can be broken in epitaxial thin films. [9] For example, orthorhombic perovskite substrates such as $DyScO_3$ (DSO) and $GdScO_3$ impose anisotropic biaxial strain on LFO, where the strain is compressive in one axis, but tensile in the other. [5] LFO films strained on (101)-oriented DSO experiences +2.94 % tensile strain along one in-plane axis and -0.22 % compression along the perpendicular axis. This strain engineering results in lowering of the symmetry from orthorhombic to monoclinic and forces the structural and AF domain from usually polydomain configuration to a single domain.

The combination of different perovskites in thin film heterostructures, e.g., ferromagnetic LSMO and antiferromagnetic LFO, adds another opportunity to control the AF order parameter. X-ray absorption spectroscopy measurements reveal that near an LSMO/LFO interface, the Fe magnetic moments in LFO become canted, and produce a net interfacial magnetisation in the LFO layer. [10] This magnetic reconstruction is due to the oxygen-octahedral connectivity, where perovskite tilt patterns propagate across the heterointerface (so-called octahedral coupling). In LSMO/LFO superlattices, octahedral coupling can alter the Mn–O–Mn and Fe–O–Fe bond angles, thereby changing the local magnetic anisotropy and the Curie temperature. [11] In this way, the interface in heterostructures provides a leverage to tune AF spin order by modifying structural distortions and exchange paths across the layers.

In this study, we investigate the structural and magnetic properties of a (La$_{0.7}$Sr$_{0.3}$MnO$_3$ /LaFeO$_3$)$_4$ superlattice grown on a (101)$_o$ DSO substrate, which imposes highly anisotropic in-plane strain. A combination of high-resolution X-ray diffraction (XRD), reciprocal space mapping (RSM), scanning precession electron diffraction (SPED) [12], and higher-order Laue zone (HOLZ) analysis were used to probe the strain relaxation mechanisms and structural domain formation. [13] Our findings reveal that the LFO layers undergo anisotropic strain relaxation, forming structural domains that propagate throughout the superlattice, while LSMO remains partially strained. Additionally, X-ray magnetic circular (XMCD) and linear dichroism (XMLD) measurements show bulk-like AF behaviour. By exploring the interplay between anisotropic strain and structural relaxation, this study contributes to a broader understanding of strain-driven effects in complex oxide superlattices.

## 2. Results and Discussion

### 2.1 Strain Analysis

Superstructures of (LSMO/LFO)$_4$ on DSO(101)$_o$ substrates were grown by pulsed laser deposition (PLD) as schematically shown in **Figure 1a),** with growth parameters optimised from previous works [5, 7]. The strain state is illustrated in Figure 1b) for LFO on DSO substrate with a tensile strain of 2.94 % in the [010]$_o$ in-plane axis and a compressive strain of -0.22 % in the [$\bar{1}$01]$_o$ perpendicular in-plane axis. For LSMO on DSO there is tensile strain in both directions, respectively, 4.06 % in the [010]$_o$ in-plane axis and 0.81 % in the [$\bar{1}$01]$_o$ perpendicular in-plane axis.

Symmetrical X-ray diffraction (XRD) 2theta-omega line scans around the (202)$_o$ peak of DSO shown in Figure 1c) reveals periodic superlattice peaks with Kiessig fringes in between, indicating a highly ordered crystalline thin film. The periodic lattice peak angles show a bilayer thickness of 13.5 nm and total superstructure thickness of 54 nm. The simulated superlattice line scan (Figure 1c)) using InteractiveXRDFit [14] yields 13 nm periodicity with LFO layer thickness of around 8.5 nm and LSMO layer thickness of 4.5 nm per bilayer. In addition, the simulation shows two distinct Kiessig fringes between the 0$^{th}$ and 1$^{st}$ order superlattice peaks. The bilayer is fitted to consist of 37 d$_{101}$ layers of LFO and 20 d$_{101}$ layers of LSMO. Rocking curves recorded of both the substrate and superlattice peaks shown in the inset of Figure 1c), reveal a FWHM of 85 arcseconds for DSO and an average of 95 arcseconds for the superlattice peaks, showing well-ordered growth.

Given that DSO imposes anisotropic strain as illustrated in Figure 1b), it is important to probe how this anisotropy impacts the thin film in the plane. High tensile strain [010]$_o$ axis is denoted as Q$_x$, in which $(424)_o^+$ and $(5\bar{1}2)_o^-$ diffraction peaks were chosen for reciprocal space maps, where the superscript '+' and '-' represent, respectively, grazing exit and grazing incidence geometry. Further, the compressive [$\bar{1}$01]$_o$ axis is denoted as Q$_y$, and the $(600)_o^+$ diffraction peak was investigated. The reciprocal space maps (Figures 1d)-f)) show vertically aligned superlattice peaks that are coherent with the observation from the 2theta-omega scan, confirming highly ordered crystal planes in Q$_z$ ([101]$_o$ out of plane) axis. Since the $(600)_o^+$ reciprocal space map in Figure 1e) shows vertically aligned superlattice peaks with the substrate peak in the $Q_y$ direction, the films are strained in the $[\bar{1}01]_o$ axis. However, in the $Q_x$ direction showing the $(424)_o^+$ and $(5\bar{1}2)_o^-$ diffraction peaks (Figures 1d) and 1f)), the superlattice peaks do not align with respect to the substrate. Hence, relaxation of the epilayers is present in the $Q_x$ direction.

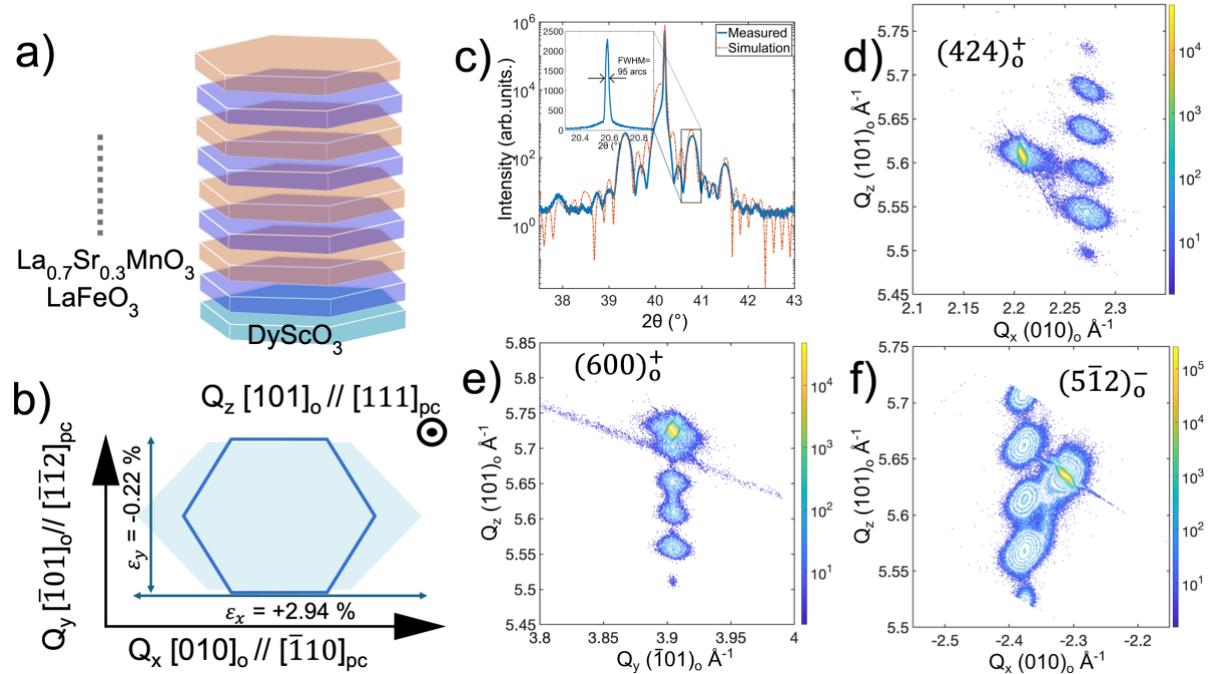

**Figure 1.** a) Schematic of the superlattice with stacking order, with LFO in blue and LSMO in orange colour. b) Illustration of the lattice mismatch in $(101)_o$ orientation between DSO (light blue) and LFO (solid blue line). $Q_{x,y,z}$ orientations are given in orthorhombic and pseuduocubic notations. c) 2 theta-omega line scan and simulation of symmetrical $(202)_o$ peak with inset showing rocking curve of the selected superlattice peak with FWHM = 95 arcseconds. d-f) Reciprocal space maps of, respectively, $(424)_o^+$, $(600)_o^+$ and $(5\bar{1}2)_o^-$ reflections.

The superlattice peaks for $Q_x$ in Figure 1d) are broad with horizontal width of 0.0158 r.l.u (relative lattice unit) for the $0^{th}$ peak versus 0.0083 r.l.u for the substrate peak. This broadening of superlattice peaks might arise from mosaicity or structural domains in the epitaxial thin films. Mosaicity in thin films are usually due to disorder of the epilayers which should be visible from rocking curves resulting in large FWHM. No observed broadening in the rocking curves in superlattice peaks in inset of Figure 1c) suggest that the broad superlattice peaks from the reciprocal space maps are due to structural domains.

Average lattice parameters from the reciprocal space maps in Figures 1d)-f) were calculated to be a=5.54 nm, b=5.58 nm and c=7.83 nm assuming a fully orthorhombic system even though LFO grown with large lattice mismatch is simulated to have monoclinic or triclinic structure. [5, 15] The estimated superlattice parameters are similar to LFO bulk values (a=5.58 nm, b=5.57 nm and c=7.83 nm [16]), which suggests that relaxation is governed by the LFO layers and LSMO is strained to LFO. The exact structure of each layer is hard to determine as the superlattice peak averages the two layers in X-ray diffraction.

The RHEED diffraction patterns in **Figures 2a)-d)** show several changes during the deposition process. The substrate had a semi-circle pattern typical of a single crystal with atomically smooth surfaces (Figure 2a)), the pattern is slightly asymmetric due to alignment. The diffraction pattern remained unchanged in position with oscillating intensities during the growth of the initial LFO layer (Figure 2b)). The recorded RHEED intensity shows 16 oscillations throughout the first layer, while the simulation from X-ray diffraction in Figure 1c) yields in average 37 $d_{101}$ layers. Thus, each recorded intensity oscillation corresponds to approximately two $d_{101}$ LFO layers deposited during growth.

At the onset of LSMO growth (Figure 2c)), a horizontal displacement in pixel position of the top spot was observed within the span of seconds. In the higher magnification inset in Figure 2c), the RHEED spot has been replaced from right (green circle) to left (blue circle) showing a rapid change in lattice spacing, suggesting relaxation to occur. However, no observed change in distance between the diffraction spots, shows no in-plane lattice spacing changes. The continuous transition in diffraction pattern from a purely semi-circle to a mixture of transmission and 2D type indicates a change in growth mode from 2D layer-by-layer to partial 3D island growth.

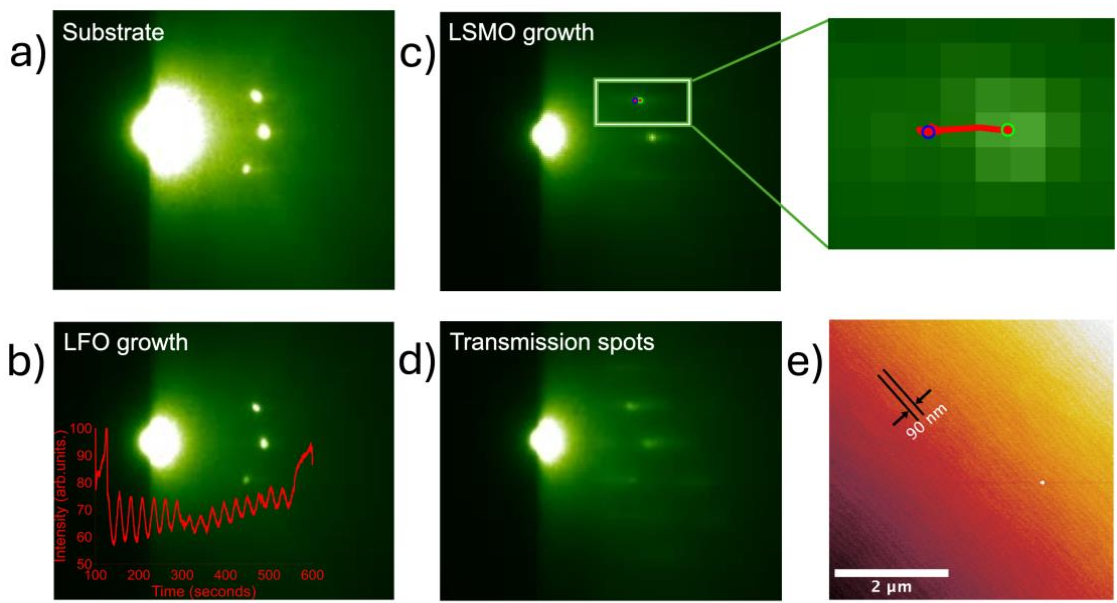

**Figure 2.** a) RHEED pattern of substrate. b) RHEED pattern with overlay showing the intensity oscillation during LFO layer growth. c) RHEED pattern from initiation of LSMO growth with a horizontal displacement of the side spot on top. Higher magnification inset shows the movement trajectory in a red line with green circle as start and blue circle as the end positions. d) Transition RHEED pattern of LSMO growth. e) AFM image of the superstructure topography with size of step-terraces shown by arrows, with the step width following the substrate.

The AFM topography image of the superlattice in Figure 2e) shows relatively smooth surface with mean roughness of 0.28 nm and clear step-and-terraces mimicking the substrate (**Figure**

**S1** in **Supplementary Information**). The AFM image indicates that the growth has not transitioned to large islands growth, but probably smaller islands with 2-3 layers were grown at once resulting in the transmission RHEED pattern. A mixed surface termination of the DSO substrate after preparation as reported by Biegalski et al. might contribute to this growth behaviour. [17, 18] The mixed termination is transferred from the substrate to subsequent growth surfaces without no annealing step between each layer growth, impacting the highly surface sensitive LSMO growth and causing islands to form. As the islands coalesce, there is an increased likelihood of strain relaxation in the unfavourable energy $[010]_o$ axis. This is coherent with the reciprocal space maps in Figures 1 d)-f) showing strain relaxation along the tensile strain $Q_x$ axis. The structural analysis by XRD in Figure 1 lacks specificity on individual layer level due to averaging nature of the in-house XRD instrument. However, the RHEED recording shows a change in diffraction pattern during the first LSMO layer growth, which call for in-depth localised analysis of the structure.

Segmented scanning precession electron diffraction (S-SPED) with the beam parallel to the $[\bar{1}01]_o$ zone axis (low strain axis) yielding in- and out-of-plane information was employed to analyse local structural strain. [12] From the S-SPED analysis shown in **Figure 3**, a thickness of 56 nm was measured for the superstructure which is in line with XRD data, with each bilayer to be around 13-14 nm. The in-plane strain analysis done by measuring the Friedel pair distance (lattice parameter) between $[040]_o$ and $[0\bar{4}0]_o$ Friedel pair (**Figure S2-SI**) resulted in the heat map in Figure 3a). The substrate and first LFO layer have approximately the same Friedel pair distance (Figure 3c)), showing that the first LFO layer was coherently strained to the substrate. A significant increase in the Friedel pair distance in the first LSMO layer was observed and can be interpreted as in-plane lattice parameter variation, showing the starting point of relaxation. This observation is coherent with the RHEED data in Figure 2 and supports that relaxation occurred first when LSMO was deposited. The following LFO and LSMO layers have almost constant in-plane distances with values around the bulk LFO lattice parameters [16] as seen in Figure 3c). Hence no further relaxation occurs, suggesting that the growth front after the first bilayer was strained to the LFO layer.

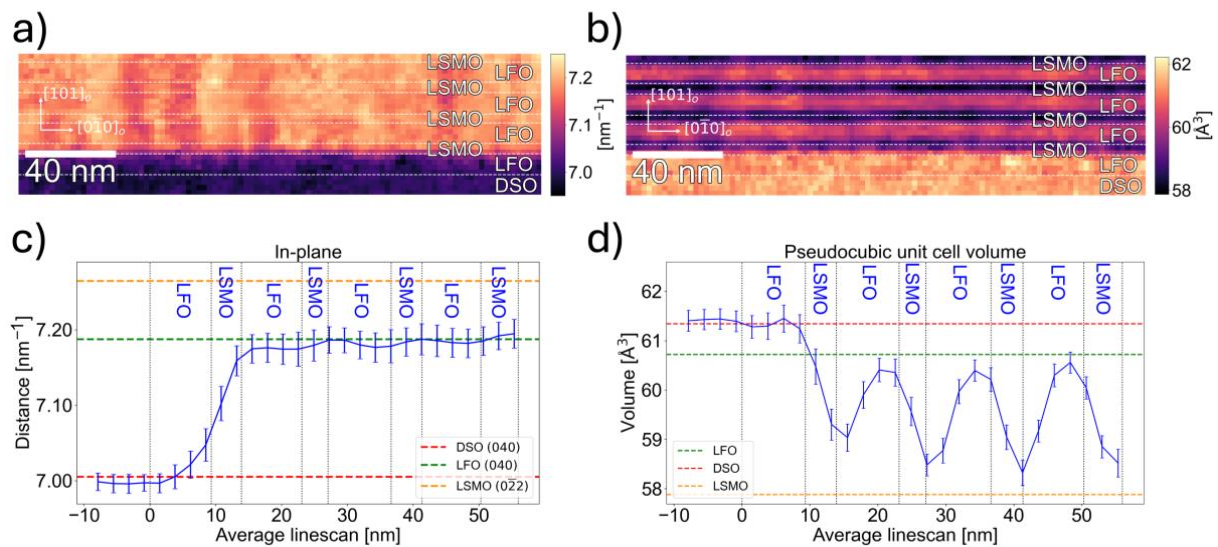

**Figure 3.** a) Heat map of Friedel pair distance between $[040]_o$ and $[0\bar{4}0]_o$ Friedel pair, probing in-plane strain in the $(010)_o$ axis of the $(LSMO/LFO)_4$ superlattice. b) Heat map of pseudocubic unit cell volume based on in- and out-of-plane distances. c) In-plane distances for $[040]$ and $[0\bar{4}0]$ Friedel pair with guidelines for DSO, LFO and LSMO lattice parameters, showing the distance converges towards LFO bulk value [16, 19, 20]. d) Pseudocubic unit cell volume calculated from lattice parameters obtained from the strain analysis. Guidelines are bulk unit cell volumes for DSO, LFO and LSMO. [16, 19, 20]

Heat map of the unit cell volume generated by using lattice parameters measured for each axis (**Figure S3-SI**) is shown in Figure 3b). The unit cell volume oscillates between the layers. Electron energy loss spectroscopy (EELS) in **Figure S4-SI** does not show chemical changes within the layers, thus the volume variation is of structural origin. The variation in unit cell volume across the superlattice is shown in Figure 3d, in which the first LFO layer has the same unit cell volume as the substrate, with a transition to a lower unit cell volume in the first LSMO layer in conjunction to the observed decrease in in-plane $[010]_o$ lattice parameter. The subsequent layers show oscillating unit cell volumes that approaches the respective bulk values due to strain relaxation with offset towards LFO bulk value. This suggests that LSMO did not fully relax to its bulk state but is strained towards the LFO lattice parameters.

Strain analysis performed in the perpendicular zone axis relative to data in Figure 3, probing the $[\bar{1}01]$ zone axis with compressive strain, showed no relaxation for the in-plane parameters (**Figure S5-SI**) which is consistent with the XRD data in Figure 1e). Thus, we can conclude the thin film underwent a partial relaxation of the high-strain axis during the first LSMO layer, and subsequent layers were strained to the second LFO layer.

## 2.2 Structural domains

The crystal structure of the superlattice was further studied by scanning transmission electron microscopy – high angle annular dark field (STEM-HAADF) as shown in **Figure 4a)**. Continuous lattice planes with clear interfaces between the layers were observed with no apparent dislocations or other structural defects in the strained $[010]_o$ zone axis. In addition, EELS in Figure S4-SI shows no significant intermixing of B-site cations at the interfaces. The STEM data in Figure 4a) shows highly crystalline lattices in the two spatial dimensions perpendicular to the electron beam, but information of the crystal structure parallel to the electron beam is limited. STEM-HOLZ technique that overcomes such spatial limitation was employed to investigate the three-dimensional crystal structure. STEM-HOLZ has previously been used in this way to analyse A-cation ordering and oxygen octahedral tilting in complex oxide thin films. [15, 21]

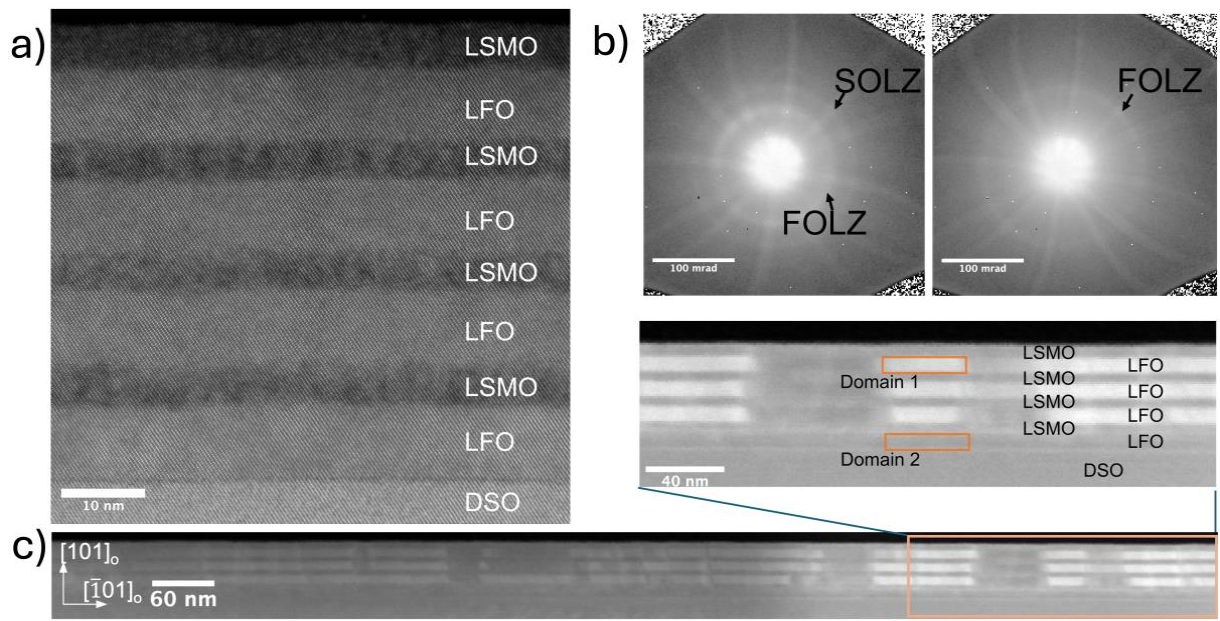

**Figure 4.** a) STEM-HAADF of the (LSMO/LFO)$_4$ superlattice. b) STEM-HOLZ scattering rings from domain 1 on the left and domain 2 on the right. c) Structural domain in 1 μm field of view and zoomed in to 200 nm. Domain 1 is highlighted by selected FOLZ ring in domain 1 as default intensity. Domain 2 has dark appearance due to absence of selected domain 1 FOLZ ring. In addition, a thickness contrast within the lamella is present from FIB fabrication giving a brighter contrast on the right hand side in the image.

Scanning transmission electron microscopy - higher order Laue zone diffraction, STEM-HOLZ results in Figure 4 b)-c) show contrasts due to two groups of HOLZ scattering rings defined as domains 1 and 2. These diffraction rings correspond to reflections of the lattice plane along the beam and represents the lattice stacking order. The difference in contrast in HOLZ observed can be attributed to the domain formation minimising the overall strain energy, avoiding structural defects such as misfit dislocations. Domain 1 consists of two rings that are the first (FOLZ) and second (SOLZ) order Laue zone scattering, whereas domain 2 only has one single outer ring (Figure 4b). To distinguish these two domains, the FOLZ ring in domain 1 was selected as the default contrast intensity. The presence of the two domains is seen by scanning across the lamella of the superstructure in 1 μm field of view and zoomed in to 200 nm as presented in Figure 4c). The substrate shows homogeneous contrast as expected from a single crystal. The first bilayer shows similar homogeneity which is consistent with a strained LFO and partially relaxed LSMO. A horizontal contrast difference is visible in the LFO layers starting from the second bilayer, confirming the presence of two different types of unit cell ordering (domains) within each layer of LFO. This suggests that the strain relaxation occurring in the first LSMO layer introduces possible structural degeneracy in LFO. The domain contrast patterns have defined vertical boundaries as observed in Figure 4c). The domain width is in the range from 40 to 100 nm. The domain width range is in good agreement with the step width of the DSO substrates (Figure S1-SI). This observation suggests that the domains nucleate from step edges during growth.

LSMO/LFO heterostructures on (111)SrTiO$_3$ substrate reported by Nord et al. show LFO with six-fold degenerate growth orientation resulting in structural twin domains in LFO depending on growth orientation.[15] Similar diffraction rings and domain contrasts as in Figure 4c) were reported, interpreted as unit cell doubling from the structural twin domains. This suggests that the HOLZ scattering contrast observed in our LMSO/LFO/DSO system also represents structural domains with a doubling of the unit cell periodicity in the LFO layers. The area in the lamella with dimmed contrast due to larger thickness, makes it difficult to quantify domain width in that area. The schematic in **Figure 5a)** illustrates three possible structural orientations of LFO and represents the two types of domains shown in STEM-HOLZ data in Figure 4. The bulk LFO $a_o$ axis is illustrated to be parallel to the DSO (101)$_o$ growth surface as seen in Figure 1b).

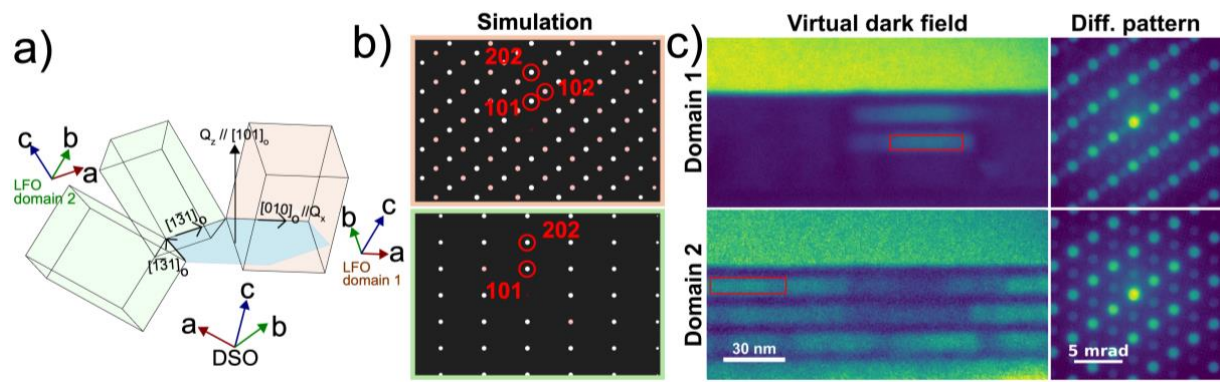

**Figure 5.** a) Schematic representation of the two possible LFO domain orientations with growth axes annotated with respect to DSO substrate. Domain 2 has a structural degeneracy. b) Diffraction simulation of the two domains in LFO by Recipro. [22] Top and bottom represent, respectively, domain 1 and 2. c) Virtual dark field image reconstructed from S-SPED data by selecting super reflection exclusively to the respective domain, highlighting the domain location throughout the superlattice.

The virtual dark field images acquired by S-SPED with simulated diffraction patterns as presented in Figure 5b-c) were used to understand the structural symmetries of each domain. Figure 5b) shows simulated diffraction patterns of LFO on DSO (101)$_o$ growth surface. The top diffraction pattern is defined as domain 1 with LFO $a_o$ axis parallel to [010]$_o$ DSO axis. The periodicity doubling in the diffraction pattern along the [001]$_o$ direction, showing the unit cell doubling as discussed above. [15] The bottom diffraction pattern is defined as domain 2 with LFO $a_o$ axis parallel to either $[\bar{1}31]_o$ or $[1\bar{3}1]_o$. These two are structurally degenerated and yields indistinguishable diffraction patterns. In addition, each structural domain can have a 90-degree rotation along the c axis resulting in structural twins for each geometry that are indifferent in terms of diffraction. Therefore, it is possible to have $b_o$ parallel to these growth axes.

Virtual dark field (VDF) images in Figure 5c) highlight the domain locations within the lamella by applying virtual apertures. The contrasts are provided by selecting super reflections that are exclusive for each domain configuration. In domain 1, diffraction points from a periodic doubling in the [001]$_o$ direction are chosen, resulting in bright regions with the rest of the lamella being dark. Conversely, domain 2 is highlighted by choosing $[\bar{1}0\bar{1}]_o$ diffraction point. The observed domains are in agreement with STEM-HOLZ data (Figure 4c) and shows clear presence of two distinctive structural domains in the superlattice. The domains in the LFO layers were translated vertically through the superstructure as also observed in STEM-HOLZ data in Figure 4b). Zhu et al. observed similar type of structural domain walls in thin LFO film grown on (100) SrTiO$_3$, which is in agreement with the results by Nord et al. [23, 15] Further, Zhu et al. showed that strained LFO grown on (001) DyScO$_3$ lifts the structural degeneracy and results in structural monodomain. The first LFO layer closest to substrate in the present superlattice is of the same structural monodomain nature, however, polydomain was observed in the superlattice after relaxation occurred in LSMO layer. The LSMO layers did not show any structural domain contrast in VDF reconstruction (**Figure S6-SI**). This type of layer selective domain that translates across layers with vertical domain walls (Figure 4c) and 5c)) has not been reported to the authors knowledge.

From the TEM data in Figures 4 and 5, strain and structural domain formation are clearly connected. The first LFO layer with fully strained configuration did not have any structural domains. Strain relaxation was observed in the first LSMO layer as seen from RHEED in Figure 2c) and strain analysis from S-SPED in Figure 3a). This establishes a basis for partial relaxation in subsequent LFO layers and the possibility for domain growth with nucleation from step edges of the substrate.

## 2.3 Magnetic properties

LSMO and LFO have magnetism closely tied to the structural arrangement, hence it is interesting to investigate how the structural domains and relaxation impact the magnetic order of each material. X-ray absorption spectra (XAS) in total electron yield mode (TEY) on top, and magnetic dichroism spectra based on the absorption data at the bottom for LSMO and LFO are provided in **Figures 6a)-b)**. The X-ray magnetic circular dichroism (XMCD) asymmetry at the bottom part of Figure 6a) shows a prototypical in-plane LSMO magnetic signature as previously reported, but with significantly lower signal magnitude. [10] In addition, the L3 peak shape in the XAS indicate a reduction of Mn ions on the surface due to surface sensitivity of the TEY measurement. [24] This reduction might be the reason for the reduced magnetisation in LSMO observed in the XMCD. The complimentary luminescence yield spectra (**Figure S7 - SI**) show correct stoichiometry with 2/3 to 1/3 ratio for Mn$^{3+/4+}$ for LSMO in the rest of the superlattice with XMCD asymmetry around 15 %, which is more as expected for 4 nm thick films of LSMO. The X-ray magnetic linear dichroism (XMLD) spectrum acquired at grazing incidence along the LFO hard axis $[010]_o$ at 80 K in Figure 6b) (bottom) shows an in-plane AF axis as previously reported, with a peak shape that indicate AF polydomain structure. [5, 25, 26] The results lead to the conclusion that structural and antiferromagnetic domains are concurrent, meanwhile Kjærness et al. reports structural monodomain results AF monodomain. [5]

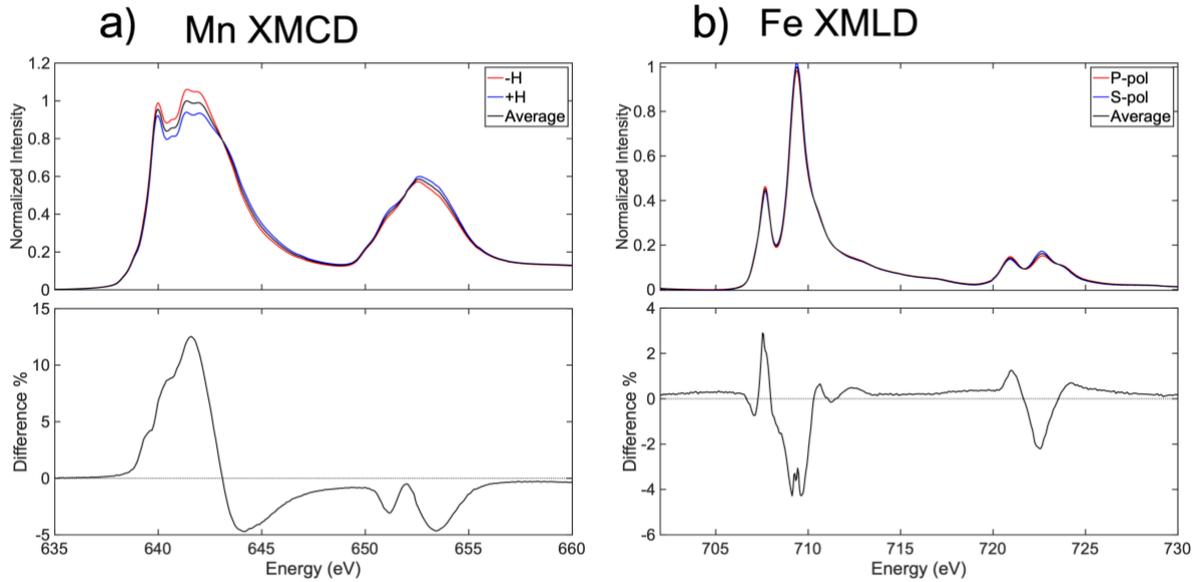

**Figure 6.** a) a) Mn X-ray absorption, with two polarizations in red and blue, with the average XAS in black on top and the resulting difference spectrum between the two polarizations at the bottom, b) Fe X-ray absorption spectrum with linear polarisation (p-polarization in red, and s-polarization in blue) and resulting difference spectrum. All data is taken at 80 K in total electron yield.

## 2.4 Outlook

Figure 6 and previous reports [5, 27] show that the structural and antiferromagnetic domains have a symbiotic relationship in LFO/LSMO superstructures, as tensile strained LFO can be either monodomain or polydomain depending on the strain state of the film to substrate. In-plane uniaxial AF monodomain is achieved when LFO layers are fully strained to DSO and polydomain is formed when LFO is in a structural degenerate state. Thus, by tuning the strain state layer by layer, one can switch between a uniaxial monodomain configuration and a polydomain state. This deterministic AF domain control via strain could be a powerful tool which can be utilized to "write" the antiferromagnetic order during growth.

The structural domains formed under anisotropic strain may also have implications for AF domain wall engineering. Previous work has shown that strain can strongly influence spin textures at domain walls (e.g., NiO on mismatched substrates [28]). By analogy, it is plausible that the domain boundaries in our superstructures affect local spin configurations which could serve as templates for engineered domain-wall behavior. Such walls, if present and being controllable, could act as channels for spin transport or be utilized in devices based on resistance changes at AF boundaries.

## 3. Conclusion

This study shows the presence of anisotropic strain induced structural relaxation and domain formation in complex LSMO/LFO superlattices on (101) oriented DSO. Selective relaxation occurred along the in-plane tensile $[010]_o$ axis, while the perpendicular in-plane compressive $[\bar{1}01]_o$ axis remained fully strained. As a result, structural domains emerge within the LFO layers, starting from the second bilayer and continuing vertically throughout the film. These domains provide an energetically favorable relaxation pathway without the formation of structural defects, and their nucleation originate with step edges on the substrate surface. Strain analysis and diffraction studies show that the LSMO layers adapt to the relaxed LaFeO$_3$ structure, being strained to the LaFeO$_3$. Magnetic characterization revealed antiferromagnetic polydomain in LaFeO$_3$ with degenerate structural domains. These findings display an interplay between strain state, domain formation, and magnetic ordering. The ability to control strain-driven domain structures opens a pathway toward deterministic manipulation of antiferromagnetic configurations via epitaxial design. This insight paves the way for exploiting anisotropic strain as a functional tool in AF domain engineering and next-generation spintronic devices.

## 4. Experimental Section

**Sample growth, X-ray diffraction and AFM**

The (La$_{0.7}$Sr$_{0.3}$MnO$_3$/LaFeO$_3$)$_4$ superlattice was deposited by pulsed laser deposition (PLD) (TSST, Twente, NL) on (101)$_o$-DyScO$_3$ substrate (SurfaceNet GMBH, Germany). The substrate was pretreated with buffered HF etching and subsequent annealing at 1050 °C for 1 h in oxygen similar to previous reports. [29] The substrate temperature during growth was set to 580 °C with a ramp of 15 K/min for both heating and cooling. The O$_2$ pressure during heating and growth was 200 mTorr and 75 Torr for in-situ annealing for 1 h. The treated substrate showed a step-and-terrace surface finish with step width range from 40 to 100 nm. A KrF excimer laser (Coherent, US) (($\lambda$ =248 nm) was used to ablate stoichiometric LSMO, LFO targets with a substrate-target distance of 50 mm at 2Hz. In-situ reflective high energy electron diffraction (RHEED) (STAIB Instruments, Germany) was recorded during the growth process. Atomic force microscopy (AFM) (Bruker, Dimension Icon, US) was used for recording topographical data. The crystalline structure was investigated using four-circle, high resolution X-ray diffractometer (XRD, Bruker D9 Discover, US), with monochromatic Cu K$\alpha$1 radiation and 0.2 mm detector slits. Reciprocal space maps (RSM) were collected with position sensitive detector with 0.002° steps in steps in $\omega$. The RSM data were collected from asymmetric reflections $(424)_o^+$, $(600)_o^+$ and $(5\bar{1}2)_o^-$ as well as the symmetric $(202)_o$ reflection, with grazing exit (+) and incidence (-) geometries to account for limited signal intensity.

**TEM and lamella preparation**

The TEM lamella was fabricated with Thermo Fisher Helios G4UX focused ion beam (FIB) (US) with standard lamella preparation method. A carbon-based protection layer was deposited to protect the thin film from Ga ion contamination, which also served as alignment tool for S-SPED data analysis.

Segmented scanning precession electron diffraction (S-SPED) datasets were acquired on a JEOL JEM-2100F (Japan), in the nanobeam diffraction mode with a convergence angle of 1.2 mrad, 1 degree precession angle and 100 Hz frequency. STEM-HOLZ and atomic resolution data were acquired on a JEOL ARM-200CF. Both instruments were equipped with a MerlinEM direct electron detector and were operated at 200 kV. Post-acquisition data processing was performed using the open-source Python libraries HyperSpy [30] and pyxem. [31]

The S-S(P)ED data were acquired with a nominal probe size of 1.39 nm, which increased with the application of precession, and a step size of 0.92 nm. Virtual dark field (VDF) images were reconstructed from the sliced segments of the S-SPED scan and used to manually approximate the probe shifts based on feature movement with the assistance of the amorphous carbon layer.

The strain analysis was carried out by choosing a Friedel pair ($hkl$) and ($\overline{hkl}$) for each lattice spacing of interest. The pixel size calibration of the detector was constant throughout the region of interest. For in-plane strain analysis, the $(040)_o$ and $(0\overline{4}0)_o$ reflections of zone axis $(\overline{1}01)_o$ were used, and the $(202)_o$ and $(\overline{2}0\overline{2})_o$ reflections were used out-of-plane analysis.

**X-ray absorption**

X-ray circular and linear dichroism data were acquired at beamline 4.0.2 at Advanced Light Source, LBNL, US in total electron yield and luminescence yield mode providing both surface magnetic sensitivity and bulk chemical composition. The experiments were done in a vector magnet chamber at 80 K cooled by liquid nitrogen with X-rays at grazing incidence angle of 35° with respect to sample surface. Total electron yield measurements for Mn were carried out with ±0.3T switching magnetic field parallel to sample surface along the beam direction and right circular polarisation. Luminescence yield measurements for Mn were carried out with ± switching circular polarisation. [25, 26] Total electron yield measurements for Fe were carried out with 0.3T magnetic field parallel to sample surface along the beam direction with ± switching linear polarisation.

**Declarations:**

**Author Contributions Statement**

Y.L. wrote the main manuscript text, prepared most figures, synthesized the materials and characterized by XRD and XAS/XMCD/XMLD. E.M., R.A. and G.K. contributed to the synthesis, which was done in the lab of G.K. T.M.D, S.B and M.N. did the TEM characterization. Analysis were done in collaboration with Y.L, M.E and I.H. C.K. supported the XAS/XMCD/XMLD characterization. All authors reviewed the manuscript.

The corresponding author declare that the authors have no competing interests as defined by Nature Portfolio, or other interests that might be perceived to influence the results and/or discussion reported in this paper.

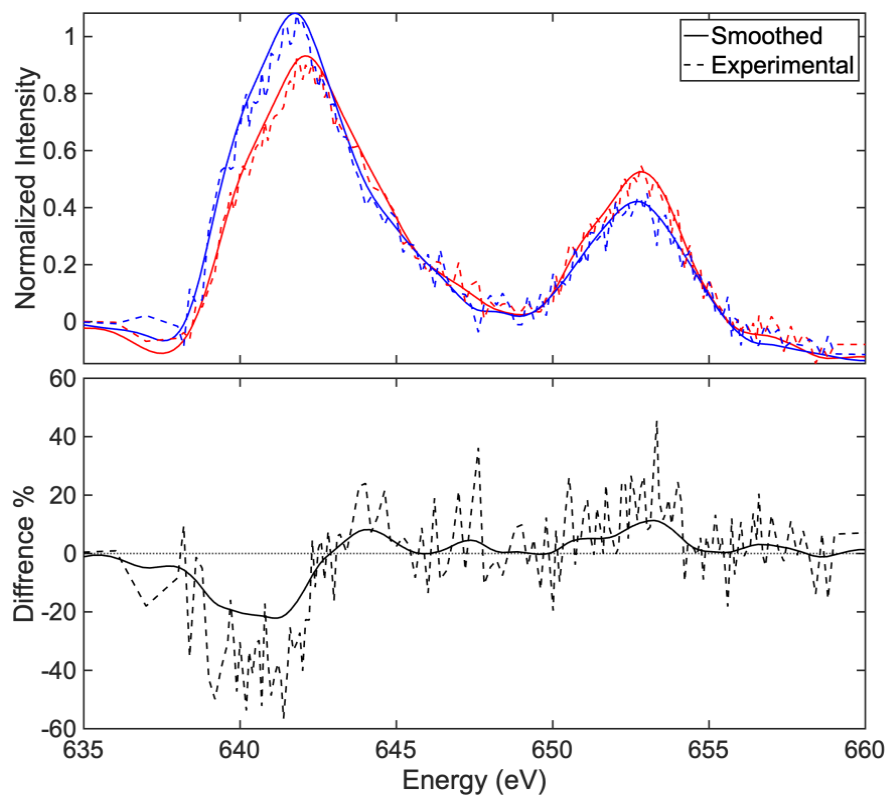

**Figure S7)** a) X-ray absorption spectra in circular polarisation for Mn energy. The observed peak splitting from reduction of $Mn^{3+}$ in TEY mode is absent, showing bulk LSMO layers retains 2/3 to 1/3 ratio for $Mn^{3+/4+}$. b) Calculated X-ray magnetic circular dichroism (XMCD) showing an asymmetry ratio around 15 % that is in range for LSMO thin films. Smoothed XMCD is shown in blue from Smooth Spline fit function.

# Supplementary Information

**Anisotropic Strain Engineering in La$_{0.7}$Sr$_{0.3}$MnO$_3$/LaFeO$_3$ Superlattice: Structural Relaxation and Domain Formation**

*Yu Liu, Thea Marie Dale, Emma van der Minne, Susanne Boucher, Romar Avila, Christoph Klewe, Gertjan Koster, Magnus Nord, Mari-Ann Einarsrud, Ingrid Hallsteinsen[*]*

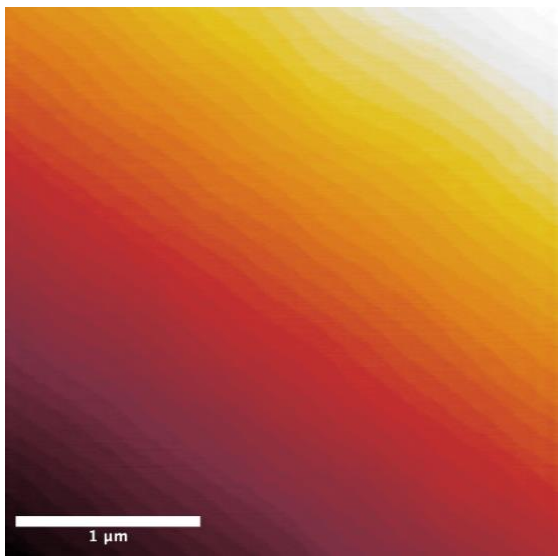

**Figure S1)** Atomic force microscopy image of the DyScO$_3$ substrate showing clear step-and-terrace topography after surface treatment.

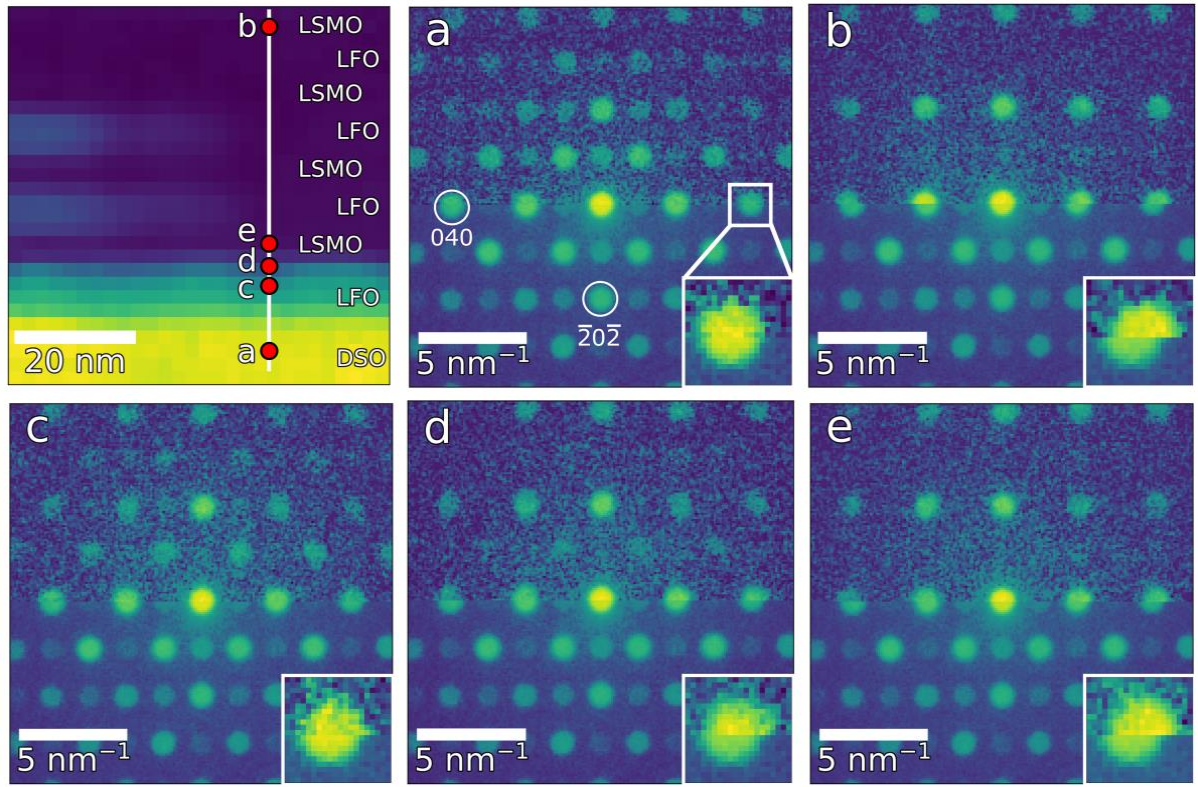

**Figure S2)** In-plane strain analysis of the [040]//[0$\bar{4}$0] Friedel pair with substrate distance as reference. Line profile with markings showing regions investigated. a) Reference distance of DSO substrate. b) Topmost layer of LSMO (upper part) compared with DSO (lower part) showing a significant displacement (zoom in of [0$\bar{4}$0]) in diffraction point due to relaxation. c) First layer of LFO (upper) compared with DSO (lower), insignificant pixel shift is seen which is in the tolerance margin. d) Beginning of the first LSMO layer (upper) compared with DSO (lower) showing significant shift in position from c), suggests a relaxation has occurred at this layer. e) End of the first LSMO layer (upper) compared with DSO (lower) showing an increased position shift from d) due to extended relaxation. However, the position is compared to b) which suggests the degree (magnitude) of relaxation is stabilised after this point for the remainder of the superlattice.

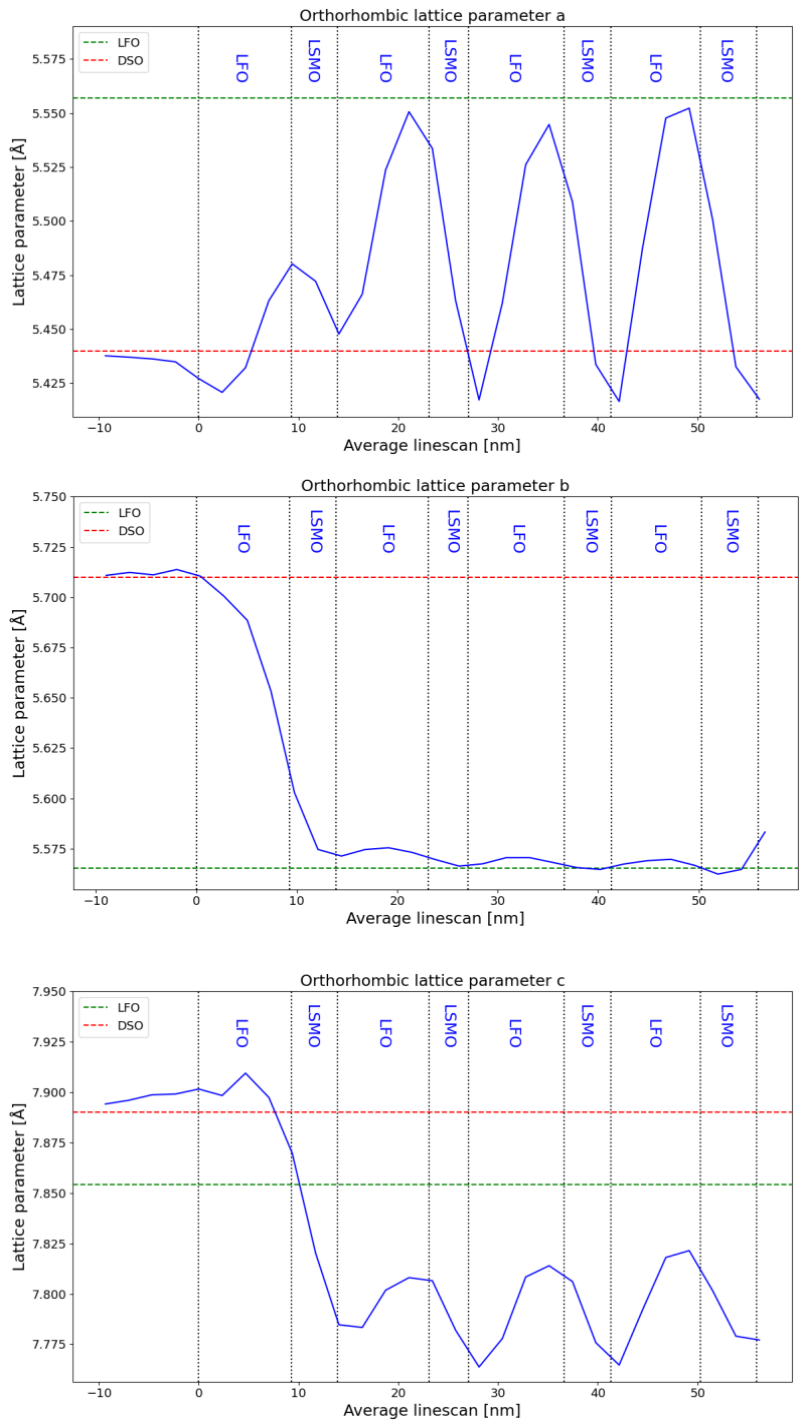

**Figure S3)** Measured orthorhombic lattice parameter a, b, c from Friedel pair diffraction points [040]//[0$\bar{4}$0] form [$\bar{1}$01] zone axis , [202]//[$\bar{2}$0$\bar{2}$] and [026]//[0$\bar{2}$$\bar{6}$] from [13$\bar{1}$] zone axis. The lattice parameter is half of the separation in distance between the diffraction pair.

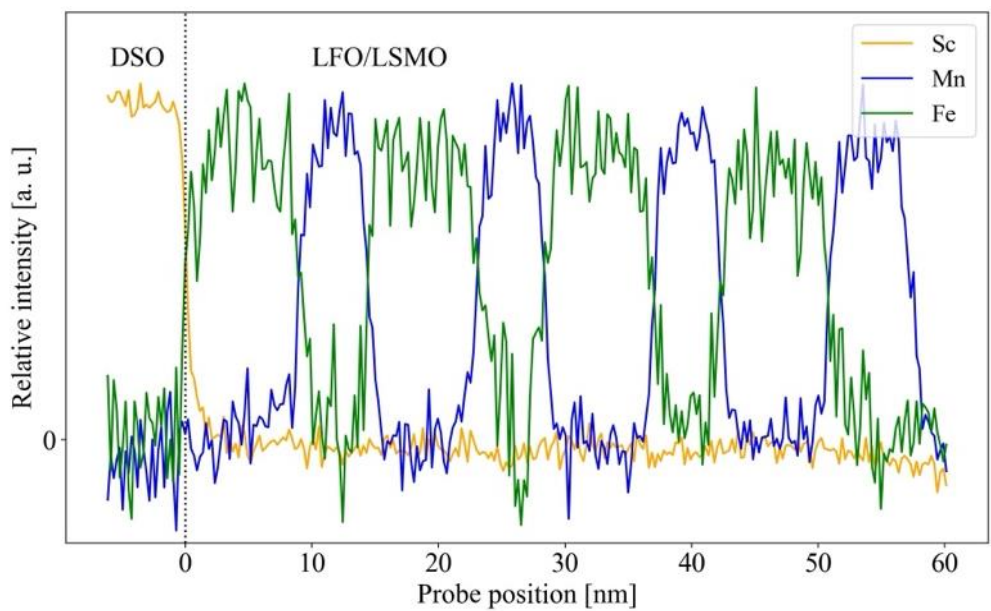

**Figure S4)** Scanning transmission electron microscopy - Electron energy loss spectroscopy line scan across the superlattice and substrate. Integrated intensities of the B-site cation edges (Sc, Mn and Fe), giving the relative element content.

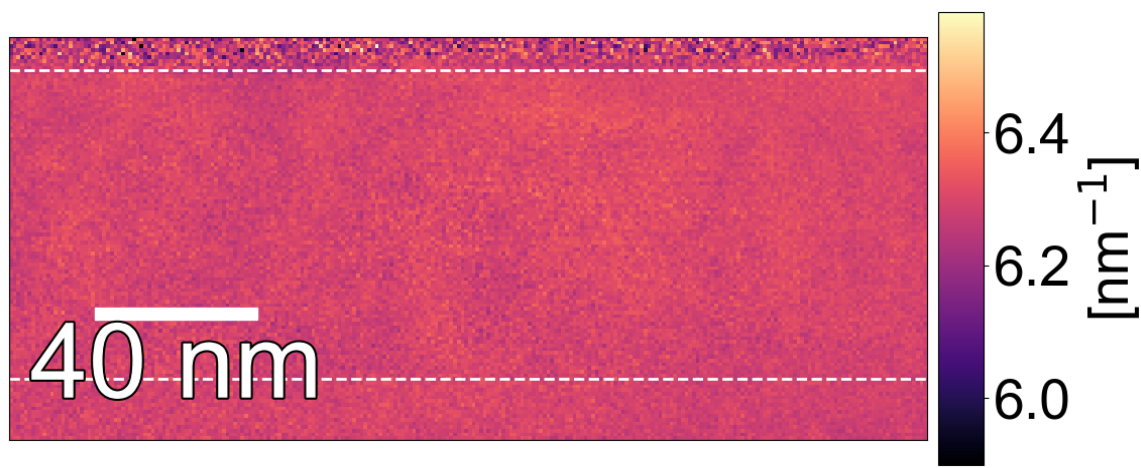

**Figure S5)** Heat map from Friedel pair $[20\bar{4}]//[\bar{2}0\bar{4}]$, showing the in-plane lattice parameter variation in the $[\bar{1}01]$ zone axis with the dotted lines indicating the superlattice location. The highly uniform heat map suggests that no relaxation is present in this zone axis.

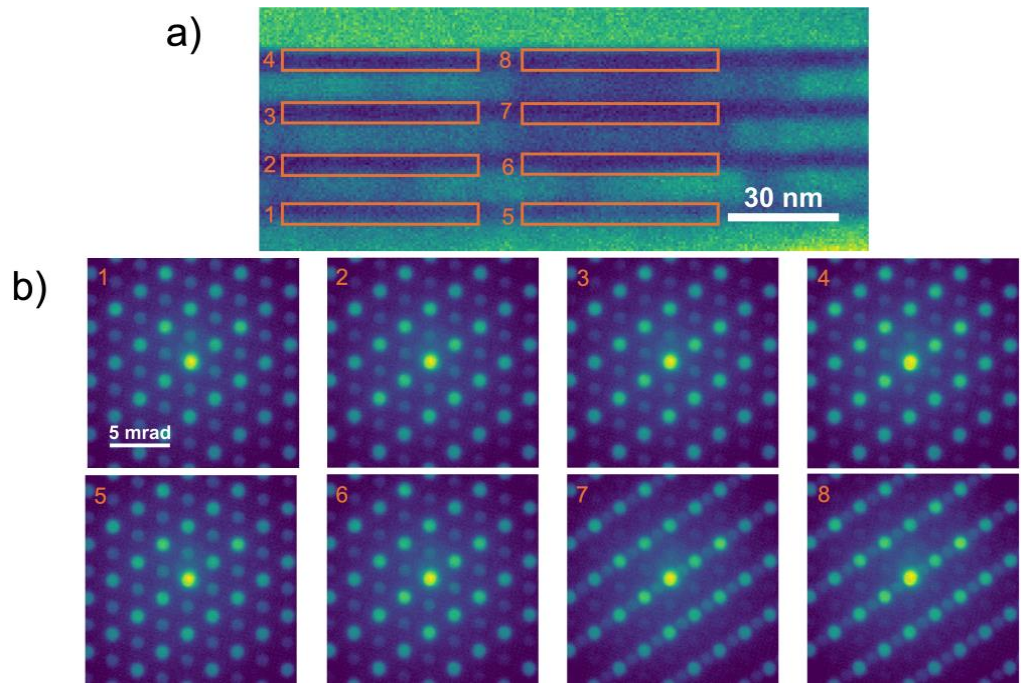

**Figure S6**) a) Virtual dark field image of the lamella with (111)$_{pc}$ virtual aperture, showing no domain contrasts within the LSMO layers. Multiple regions are selected for manual inspection b) virtual diffraction patterns from regions in VDF image above, showing no real structural changes within the layers. A faint superreflection is observed in 7-8) near LFO domain 1, which might arise from dynamic scattering effects instead of structural changes.